\documentclass[twocolumn,prl,showpacs,superscriptaddress,citeautoscript]{revtex4}
\pdfoutput=1
\usepackage{graphicx}
\usepackage{bm}

\begin{document}

\title{Magnetically asymmetric interfaces in a (LaMnO$_3$)/(SrMnO$_3$) superlattice due to structural asymmetries}

\author{S.\ J.\ May} \email{smay@anl.gov} 
\affiliation{
Materials Science Division, Argonne National Laboratory, Argonne, IL 60439\\}
\author{A.\ B.\ Shah}
\affiliation{
Department of Materials Science and Engineering, University of Illinois, Urbana-Champaign, IL 61801\\}
\author{S.\ G.\ E.\ te Velthuis}
\affiliation{
Materials Science Division, Argonne National Laboratory, Argonne, IL 60439\\}
\author{M.\ R.\ Fitzsimmons}
\affiliation{
Los Alamos National Laboratory, Los Alamos, NM 87545\\}
\author{J.\ M.\ Zuo}
\affiliation{
Department of Materials Science and Engineering, University of Illinois, Urbana-Champaign, IL 61801\\}\author{X.\ Zhai}
\author{J.\ N.\ Eckstein}
\affiliation{
Department of Physics, University of Illinois, Urbana-Champaign, IL 61801\\}
\author{S.\ D.\ Bader}
\author{A.\ Bhattacharya}
\email{anand@anl.gov}
\affiliation{
Materials Science Division, Argonne National Laboratory, Argonne, IL 60439\\}
\affiliation{
Center for Nanoscale Materials, Argonne National Laboratory, Argonne, IL 60439}

\date{\today}
             

\begin{abstract}
Polarized neutron reflectivity measurements of a ferromagnetic [(LaMnO$_3$)$_{11.8}$/(SrMnO$_3$)$_{4.4}$]$_6$ superlattice reveal a modulated magnetic structure with an enhanced magnetization at the interfaces where LaMnO$_3$ was deposited on SrMnO$_3$ (LMO/SMO).  However, the opposite interfaces (SMO/LMO) are found to have a reduced ferromagnetic moment.  The magnetic asymmetry arises from the difference in lateral structural roughness of the two interfaces observed via electron microscopy, with strong ferromagnetism present at the interfaces that are atomically smooth over tens of nanometers.  This result demonstrates that atomic-scale roughness can destabilize interfacial phases in complex oxide heterostructures.
\end{abstract}

\maketitle

   Recent advances in thin film deposition techniques permit the exploration of novel phases emerging at atomically sharp interfaces between dissimilar transition metal oxide compounds.  These interfaces can exhibit properties not present in either of the adjoined compounds, such as metallic conductivity between a band insulator (SrTiO$_3$) and a Mott insulator (LaTiO$_3$), or ferromagnetism between an antiferromagnet (CaMnO$_3$) and a paramagnet (CaRuO$_3$).\cite{Ohtomo02,Takahashi01}  The manganites, in particular, are excellent candidates in the search for novel interfacial phases due to the wide range of charge, orbital and magnetic ordering phenomena they exhibit.\cite{Salamon01}  In bulk form, LaMnO$_3$ (LMO) is a Mott insulator (Mn$^3$$^+$) with \textit{A}-type antiferromagnetic ordering, while SrMnO$_3$ (SMO) is a band insulator (Mn$^{4+}$) with \textit{G}-type antiferromagnetic ordering.  Alloying LMO and SMO yields the mixed-valence (Mn$^{3+/4+}$) compound, La$_{1-x}$Sr$_x$MnO$_3$ (LSMO), which exhibits double-exchange-mediated ferromagnetism for $0.15 < x \leq 0.5$.  Similarly, one may expect a ferromagnetic region of one or two unit cells to arise at a LMO/SMO interface due a local mixed-valence state brought about by the transfer of $e_g$ electrons from LMO into SMO.\cite{Lin06}  
   
   Investigations of (LMO)$_k$/(SMO)$_j$ superlattices ($0.2 \leq j/(k +j) \leq 0.5$), where $k$ and $j$ are the number of unit cells in each layer,  have examined their macroscopic magnetic and electronic transport properties,\cite{Salvador99,Koida02} structural and chemical profile,\cite{Verbeeck02} interfacial density of states,\cite{Satoh05,Smadici07} and the metal-insulator transition that occurs as the distance between interfaces is increased.\cite{Anand07b}  The collective results of these studies indicate that superlattices comprised of thin bilayers ($k + j \leq 8$) are ferromagnetic metals similar to bulk LSMO compounds, while superlattices with thick bilayers ($k + j \geq 9$) are insulating with reduced values of the magnetization and Curie temperature ($T_C$).  While the ferromagnetism measured in the latter class of superlattices is often assumed to reside at the interfaces, direct measurements of the magnetic structure have not been reported.  Additionally, the issue of how sensitive the interfacial ferromagnetism is to structural properties such as roughness and interlayer diffusion has yet to be explored.  

   We have investigated the magnetic structure of a [(LMO)$_{11.8}$/(SMO)$_{4.4}$]$_6$ superlattice using polarized neutron reflectivity (PNR).  Fits to the PNR data indicate the presence of an asymmetric interfacial magnetic structure that repeats every bilayer.  An enhanced ferromagnetic moment is present at the LMO/SMO interfaces but not the SMO/LMO interfaces.  The magnetic asymmetry arises from a difference in the structural roughness of the LMO/SMO and SMO/LMO interfaces.

   The superlattice was deposited on an insulating SrTiO$_3$ substrate using ozone-assisted molecular beam epitaxy.  Details of the deposition procedure are reported in Ref. ~\onlinecite{Anand07}.  X-ray reflectivity and diffraction measurements were carried out on a Philips XPert diffractometer.  PNR measurements were made using the ASTERIX instrument at LANSCE in the Los Alamos National Laboratory.\cite{Fitz05}  The footprint of the collimated neutron beam is larger than the sample (14 x 14 mm$^2$) for all angles of the measurement.  Scanning transmission electron microscopy (STEM) was performed using a JEOL 2200FS operated at 200 kV.  Details for the sample preparation are presented elsewhere.\cite{Shah08}
   
\begin{figure}
\includegraphics[width=2.1in]{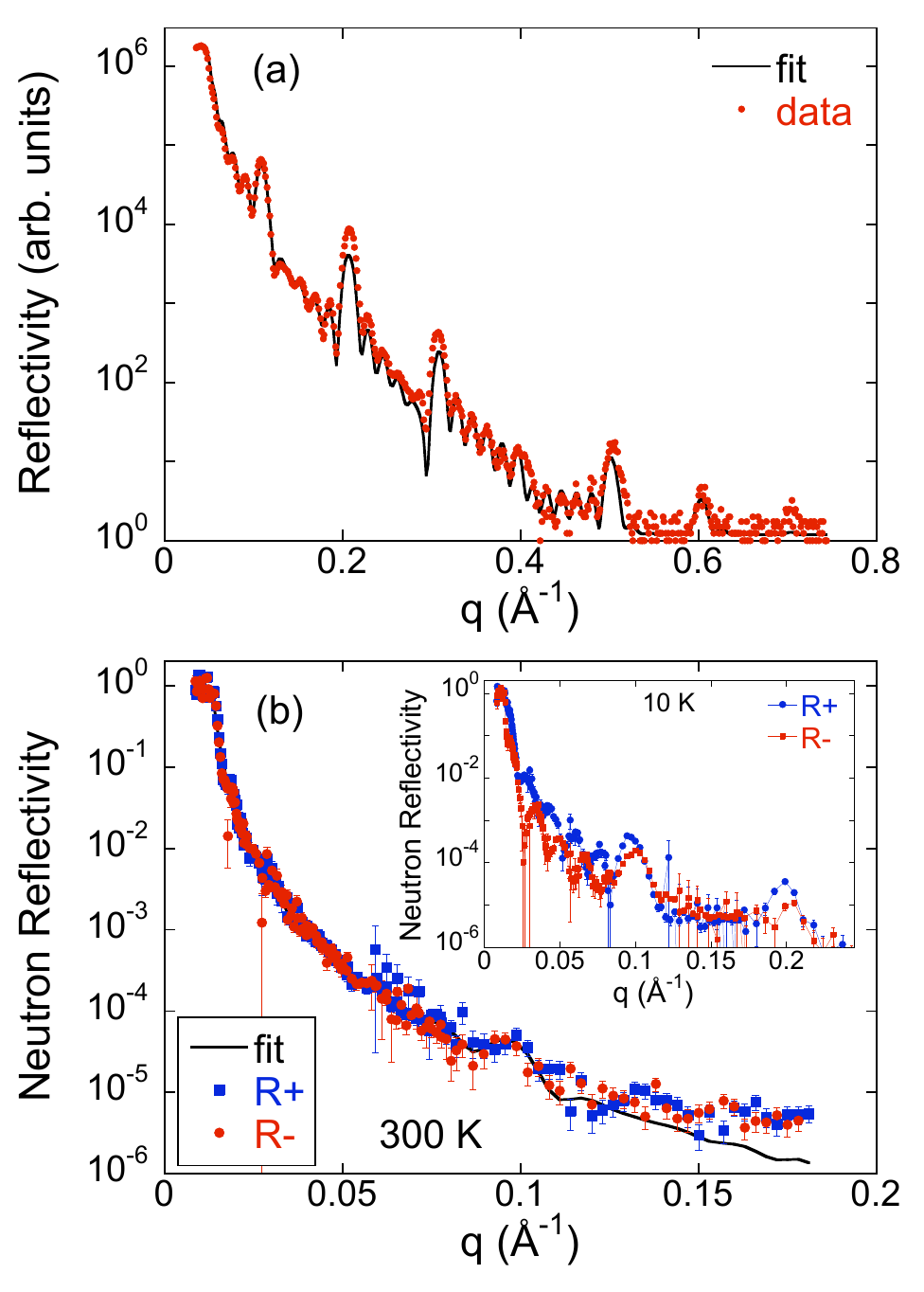}
\caption{X-ray reflectivity of the superlattice (a).  The solid line shows the fit yielding the composition.  The PNR measured at 300 K is given in (b).  The solid line shows the fit yielding the nuclear scattering length densities of LMO and SMO.  For comparison, the PNR measured at 10 K is shown in the inset of (b).}
\label{fig:structure}
\end{figure}

   The superlattice composition was determined from refinements of x-ray reflectivity data.  Thickness fringes and superlattice peaks are visible in the reflectivity data for all samples, indicating high quality interfaces persisting over macroscopic distances.  High resolution STEM images confirm the interfaces are atomically abrupt with interlayer diffusion confined to one unit cell.\cite{Smadici07}  The x-ray reflectivity data and refinement for the superlattice is shown in Fig.~\ref{fig:structure}.  The presence of a third Bragg peak ($q = 0.3$~\AA$^{-1}$), which should be suppressed if the ratio of LMO to SMO is 2:1, and the suppression of the fourth Bragg peak ($q = 0.4$~\AA$^{-1}$) indicate that the superlattices have excess LMO and are deficient of SMO in deviating from a 2:1 ratio.  The best fit to the data yield a composition of [(LMO)$_{11.8}$/(SMO)$_{4.4}$]$_6$ with the top LMO layer thinner than the other LMO layers (10.8 unit cells) and one unit cell of SMO capping the structure.\cite{Refl}  Out-of-plane lattice parameters of 3.945 and 3.714 \AA~were assumed for LMO and SMO, respectively.\cite{Anand07}  The average c-axis parameter of the superlattice is 3.874$\pm$0.003\AA, as determined from x-ray diffraction.
      
   PNR is a well-established technique used to resolve the magnetic depth profiles of heterojunctions and superlattices.\cite{Fitz05}  The scattering length density profile of the sample, which includes a nuclear and a magnetic contribution, can be extracted from the reflectivity data measured with the neutrons polarized parallel ($R^{+}$) and antiparallel ($R^{-}$) to the magnetization of the sample.  We did not perform polarization analysis, as the magnetization was assumed to be parallel to the in-plane field (0.55 T) applied throughout the measurement.  

\begin{figure*}
\includegraphics[width=6.7in]{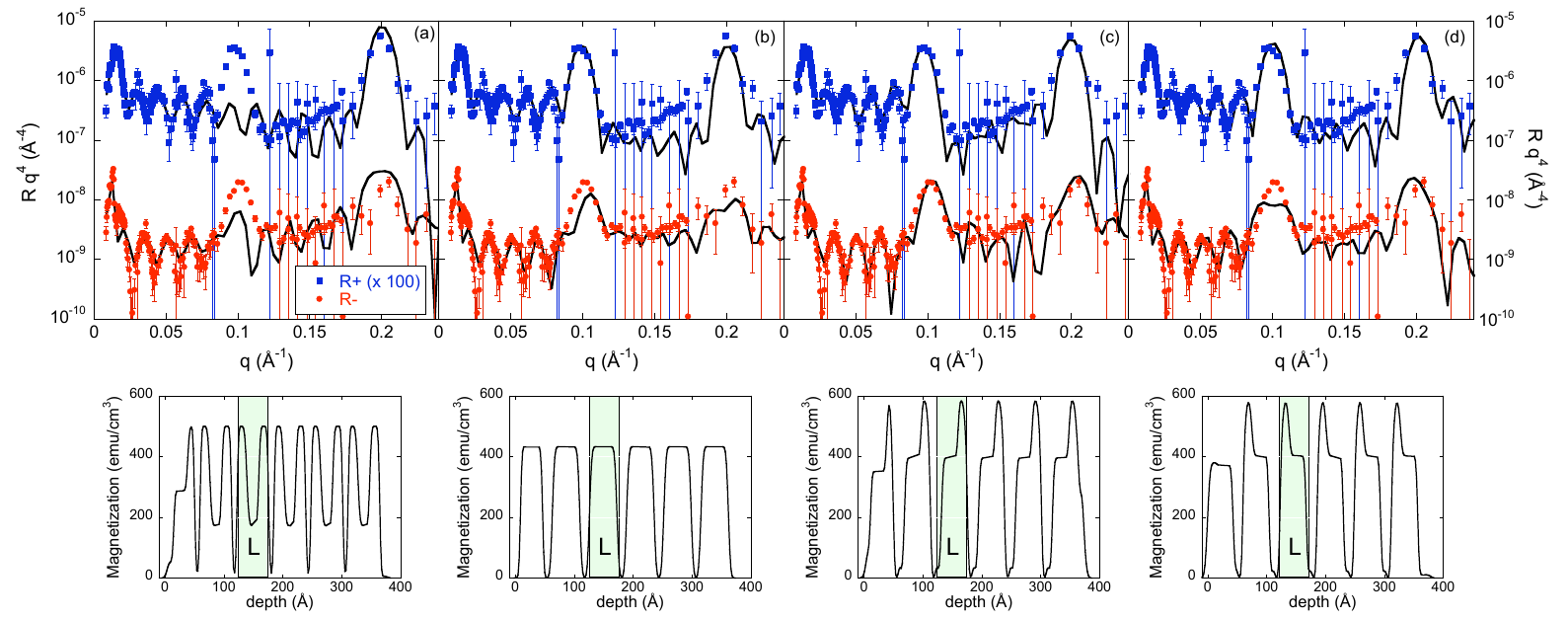}
\caption{PNR multiplied by $q^4$ measured at 10 K.  For clarity, the $R^{+}$ data has been multiplied by 100.  The error in $q$ is contained within each data point.  The black lines show fits obtained assuming (a) a large magnetization is present at both LMO/SMO and SMO/LMO interfaces with an equal magnitude, (b) a constant magnetization is present in the LMO layers with a negligible magnetization in the SMO layers, and (c,d) the magnetization is asymmetric about the LMO/SMO and SMO/LMO interfaces.  The fit in (c) is the best fit obtained for the PNR data.  The corresponding magnetic depth profile is given below each fit.  The shaded boxes (labeled L) highlight the profiles within a LMO layer.}
\label{fig:PNR}
\end{figure*}
   
   Figure~\ref{fig:structure}(b) shows the PNR data measured at 300 K, well above the Curie temperature of the superlattice ($T_{C} \approx$ 180 K).  The $R^{+}$ and $R^{-}$ reflections are equal, confirming the lack of ferromagnetism at room temperature.  The nuclear potentials of the LMO and SMO layers were determined by fitting the room temperature neutron reflectivity, assuming zero magnetization in the sample, simultaneously with the x-ray reflectivity.  The nuclear scattering length densities (SLDs) obtained from the fit are 3.75 and 3.55 x 10$^{-6}$ \AA$^{-2}$, in good agreement with the calculated values for the LMO and SMO of 3.64 and 3.65 x 10$^{-6}$ \AA$^{-2}$, respectively.  Due to the small contrast between the nuclear SLDs of LMO and SMO, the Bragg reflection at $q$ = 0.1 \AA$^{-1}$ is weak and the thickness oscillations are negligible.  The superlattice was then field-cooled in 0.55 T to 10 K.  At these conditions, the sample magnetization ($M = 1.93$ $\mu_B$/Mn) is nearly saturated ($M_{sat} = 2.02$ $\mu_B$/Mn).  The low temperature PNR data are shown in the inset of Fig.~\ref{fig:structure}(b).  A difference is observed between the $R^{+}$ and $R^{-}$ data at 10 K, indicating the sample magnetization is contributing to the scattering potential. Bragg reflections are seen at \textit{q} = 0.1 and 0.2 \AA$^{-1}$, corresponding to a modulation of the scattering potential.  Using the equation $q = 2\pi n/d$, the repeat period of the potential, $d$, is determined to be 62.6 \AA, which is equal to the thickness of each bilayer.  As demonstrated by the room temperature measurements, the modulation of the nuclear potential is insufficient to produce the Bragg peaks measured at 10 K.  Instead, a modulation of the magnetic potential that is $\sim$ 6 times larger than the difference in the nuclear potentials of SMO and LMO is required to create Bragg peaks of the same magnitude as those observed at 10 K.  Thus, the measured Bragg reflections rule out the possibility that the magnetization is constant throughout the superlattice.  
   
   Dozens of possible magnetic structures were employed as initial guesses to fit the PNR data using the co\_refine computer routine.\cite{Fitz05}  The fitting parameters were restricted to ensure that the local magnetization did not exceed 4 $\mu_B$/Mn anywhere in the structure.  The PNR and x-ray reflectivity were fit simultaneously to ensure the structural properties obtained from the neutron and x-ray data are in agreement.  In the fits, all ferromagnetic moments are assumed to be parallel to the in-plane applied field.  The magnetization in the surface layers and the SMO layers directly adjacent to the SrTiO$_3$ substrate was free to differ from the rest of the superlattice.  
   
   Figure~\ref{fig:PNR}(a) shows the fitting results obtained by constraining the magnetic potentials at the interfaces to be equal while allowing the other parameters to vary; the magnitude and thickness of the magnetic potentials were free to vary in order to optimize the fit.  This produces a magnetic profile in which ferromagnetism arises symmetrically at the interfaces while the non-interfacial layers of the superlattice are antiferromagnetic.\cite{Lin06,Koida02}  Specifically in Fig.~\ref{fig:PNR}(a), the three LMO unit cells and one SMO unit cell at each interface are ferromagnetic ($\sim$ 3.3 $\mu_B$/Mn), while the remaining SMO and LMO layers possess reduced magnetizations.  Likewise, all other fits with magnetically symmetric interfaces were unable to reproduce the data.  In general, the intensities of the first Bragg reflections ($q$ = 0.1~\AA$^{-1}$) calculated from profiles with large, symmetric interfacial magnetizations are less than what was measured by roughly an order of magnitude.  
   
   Figure~\ref{fig:PNR}(b) shows the fitting results obtained by constraining the magnetic profile to be commensurate with the chemical profile, while allowing the magnetization of the LMO and SMO layers to vary.  This set of constraints produces poor fits, ruling out the possibility that there is a constant magnetization in the LMO layers and no magnetization in the SMO layers.  Note the disagreement between the fit and the data at the $R^{-}$ Bragg reflections near $q$ = 0.1 and 0.2 \AA$^{-1}$.  
   
   Figure~\ref{fig:PNR}(c) shows the fitting results obtained by splitting each LMO and SMO layer into three sublayers, the magnetization and thickness of which were free to vary.  This method produced the best fit to the data, yielding a magnetic structure in which enhanced interfacial ferromagnetic moments (3.8 $\mu_B$/Mn) arise at each LMO/SMO junction, extending over three LMO unit cells.  Unexpectedly, a reduced ferromagnetic moment is present at the SMO/LMO interfaces with $<$ 0.1 $\mu_B$/Mn in the 6 \AA~of LMO at the interface.  The non-interfacial LMO layers have a moment of 2.6 $\mu_B$/Mn, consistent with reports of magnetization in LMO films grown STO.\cite{Anand07}\cite{Tanaka00}  In contrast, the non-interfacial SMO layers (one unit cell from the interface) have a negligible magnetization ($<$ 0.1 $\mu_B$/Mn).  The integrated magnetization obtained from the fit is within 6 $\%$ of the value measured by SQUID magnetometry.  Statistically, the fit shown in Fig.~\ref{fig:PNR}(c) is a significant improvement over the fit shown in Fig.~\ref{fig:PNR}(b).  The chi-squared values of the first $R^{-}$, second $R^{-}$, and second $R^{+}$ Bragg reflections are 2.5, 1.6, and 3.1 times larger, respectively, in the commensurate profile (Fig.~\ref{fig:PNR}(b)) than in the profile obtained from the best fit (Fig.~\ref{fig:PNR}(c)).
   
   While a symmetric profile may be expected from purely physical arguments, issues related to materials synthesis can play a role in stabilizing the interfacial ferromagnetic structure.  Local strain variations, differences in the surface roughness of the two interfaces, or atomic segregation effects\cite{Kourkoutis07} may be the origin of asymmetric interfacial properties.  The asymmetric magnetic profile is an unexpected result that illustrates the need for advanced characterization of interfacial ordering phenomena.  PNR is well suited to study interfacial ferromagnetism as it enables the separation of contributions from different interfaces.  Figure~\ref{fig:PNR}(d) illustrates this point, showing the fit obtained by forcing the enhanced (reduced) ferromagnetic moment to reside at the SMO/LMO (LMO/SMO) interface, opposite to what was determined from the best fit [Fig.~\ref{fig:PNR}(c)].  The fit shown in Fig.~\ref{fig:PNR}(d) does not reproduce the $R^{-}$ Bragg reflections.
   
\begin{figure}
\includegraphics[width=3.0in]{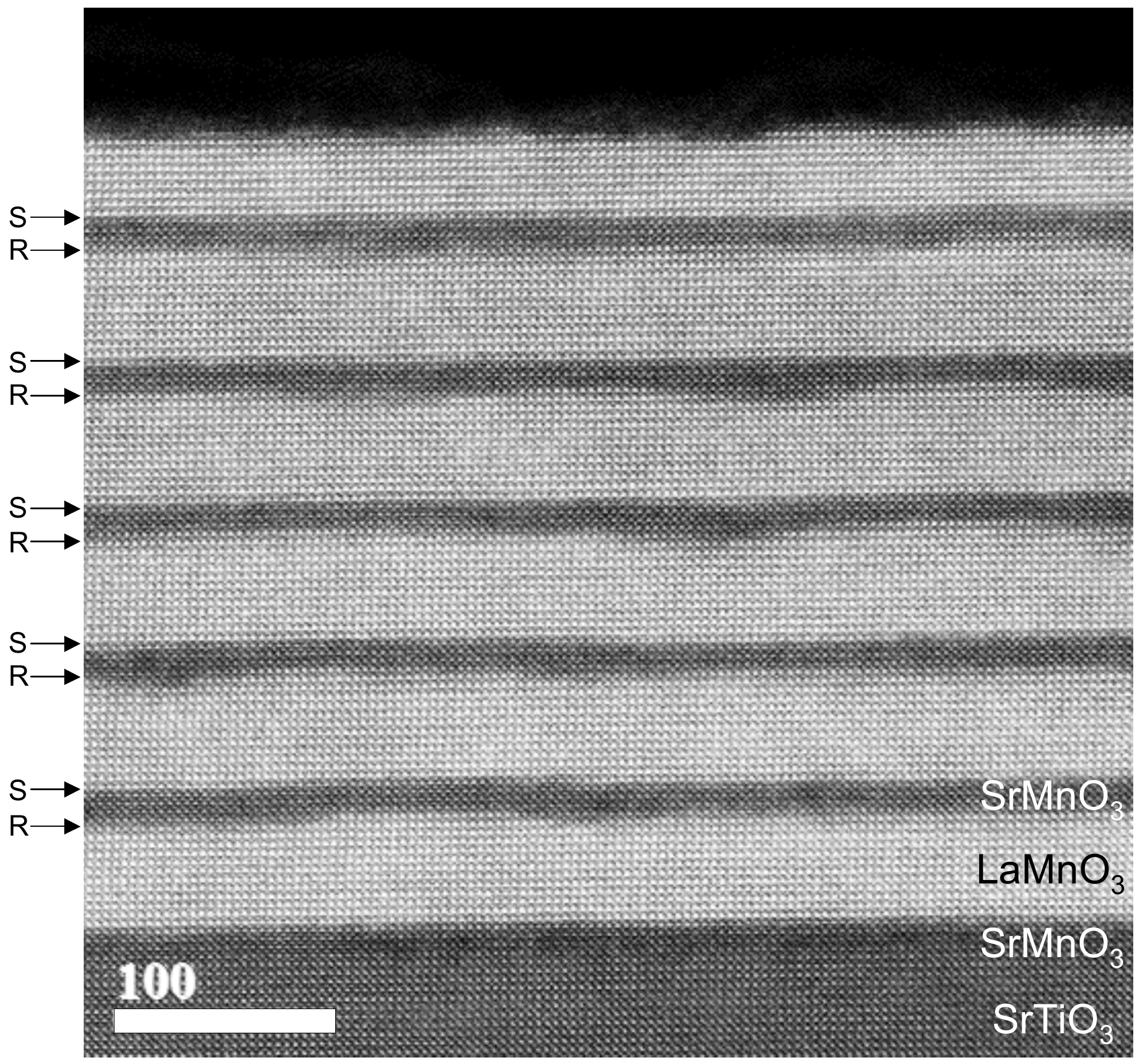}
\caption{Transmission electron microscopy image of the superlattice.  The LMO/SMO interfaces (labeled S) remain atomically flat over tens of nanometers, while the SMO/LMO interfaces (labeled R) are rough.  Units of error bar are Angstroms.}
\label{fig:TEM}
\end{figure}      

\begin{figure}
\includegraphics[width=2.7in]{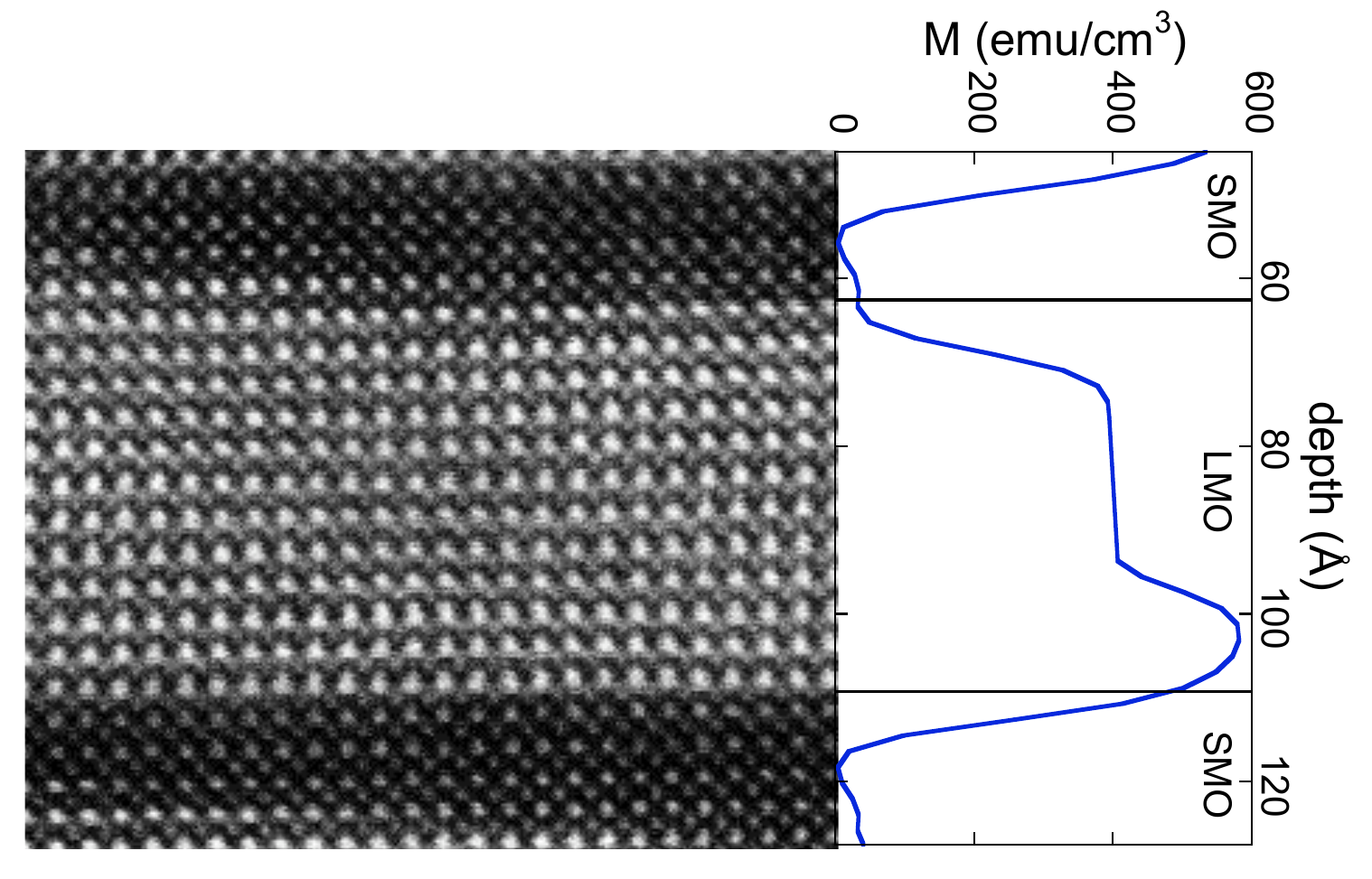}
\caption{Comparison of the structural and magnetic profile of a SMO/LMO/SMO repeat unit in the superlattice.  Note that the STEM image is a local measurement, while the magnetic profile is an average obtained over the whole sample.}
\label{fig:Profile}
\end{figure}   
   
   Z-contrast STEM was used to investigate the atomic structure of the superlattice. The interfaces were found to be structurally asymmetric, as shown in Fig.~\ref{fig:TEM}. The LMO/SMO interfaces are atomically smooth within 1 unit cell over tens of nanometers laterally.  The interfacial row of $A$-site atoms is predominately La atoms indicating the LMO/SMO interfaces are atomically abrupt within one unit cell.  The intensity ($I$) of scattering from these interfacial $A$-site atoms is equal to $0.77 \pm0.1(I_{La}-I_{Sr}) + I_{Sr}$.  The SMO/LMO interfaces are also atomically abrupt, however they consist of plateaus and valleys with a peak-to-valley height of 2 unit cells and atomic steps every 5 to 15 nm laterally in all supercells.  The SMO fills the steps and grows flat. The next LMO layer forms a sharp interface with the flat SMO, but again forms steps at the top unit cells of growth. The average thickness of the LMO and SMO layers determined from STEM are 11.3$\pm$0.6 and 4.9$\pm$0.8 unit cells (excluding the top LMO layer).  The non-integer layer thicknesses have to be accommodated at the interfaces. In the usual equilibrium view of wetting and film growth, the roughness at the interface between two materials is governed by their individual surface energies and the interface energy.\cite{Bauer86}   However, the kinetics of film growth, which depend upon the flux of atoms during growth, temperature, and morphology of the substrate, may be critical. It is possible that any of these factors lead to the LMO/SMO interfaces being smoother than their SMO/LMO counterparts.  A side-by-side comparison of the magnetic profile and a STEM image is given in Fig.~\ref{fig:Profile}, which shows large magnetic moment at the sharp LMO/SMO interface and reduced moment at the rough SMO/LMO interface. This correlation between the microstructure and the magnetic depth profile suggests the interfacial ferromagnetic phase is highly sensitive to roughness, the presence of which destabilizes the ferromagnetic order.
      
   In conclusion, we have measured the magnetic structure in a ferromagnetic [(LMO)$_{11.8}$/(SMO)$_{4.4}$]$_6$ superlattice.  Enhanced interfacial ferromagnetism was observed at each LMO/SMO interface, while the SMO/LMO interfaces have reduced ferromagnetic moments.  The magnetic asymmetry arises from the difference in roughness of the two interfaces, with strong ferromagnetism present at the interfaces that exhibit little structural roughness.  In the context of charge transfer affecting ferromagnetism,\cite{Lin06} this result seems to indicate that interfacial ferromagnetism is better realized at an atomically smooth interface and can be destabilized by structural roughness.
   
   We thank Axel Hoffmann for discussions. Work was supported by the U.S. Department of Energy, Office of Basic Energy Sciences under contracts DE-AC02-06CH11357 (at Argonne), DEFG02-91-ER45439, DE-FG02-07ER46453 and DE-FG02-07ER46471 (at the UIUC Frederick Seitz Materials Research Lab), and DE-AC52-06NA25396 (at Los Alamos National Lab's Lujan Neutron Scattering Center at LANSCE.)


\begin{thebibliography}{17}
\bibitem[{Ref()}]{Refl}
\bibinfo{note}{The reflectivity represents an average over the 1.4 x 1.4 cm$^2$
  sample. From the deposition geometry, we expect gradients of up to five
  percent in the La/Sr/Mn ratio across the substrate.}
\end{thebibliography}

\end{document}